\documentclass[12pt,twoside]{article}
\input epsf
\usepackage{psfig}
\newcommand{\VolumeHeader}{}
\newcommand{\VolumeSerial}{LNS}

\newcommand{\ActivityName}{ {\normalsize {\it 
Gravitational Waves - A Challenge to Theoretical Physics
}}\\
{\normalsize {\it 
}}}
\newcommand{\ActivityDate}{ {\normalsize {\it
Trieste, 2-7 June 2000 
}}}

\newcommand{\pa}{\partial}
\newcommand{\wt}{\hat{t}}
\newcommand{\wH}{\widehat{H}}
\newcommand{\ww}{\widehat{\omega}}
\newcommand{\wF}{\widehat{ F}}
\newcommand{\be}{\begin{equation}}
\newcommand{\ee}{\end{equation}}
\newcommand{\bea}{\begin{eqnarray}}
\newcommand{\eea}{\end{eqnarray}}

\newcommand{\LectureHeader}{Detecting Black Hole Binaries}

\begin{document}
\pagestyle{myheadings}
\markboth{\LectureHeader}{\VolumeHeader}
\markright{\VolumeHeader}


\begin{titlepage}


\title{Detecting Binary Black Holes With Efficient and Reliable Templates}

\author{T. Damour$^\dagger$, B.R. Iyer$^\ddagger$ and 
B.S. Sathyaprakash$^{\S}$\thanks{B.Sathyaprakash@CARDIFF.AC.UK}
\\[1cm]
{\normalsize
{\it $^\dagger$ IHES, Bur-Sur Yvette, France.}}
\\
{\normalsize
{\it $^\ddagger$ Raman Research Institute, Bangalore, India.}}
\\
{\normalsize
{\it $^\S$ Department of Physics and Astronomy, Cardiff University,
Cardiff, UK.}}
\\[10cm]
{\normalsize {\it Lecture given at the: }}
\\
\ActivityName 
\\
\ActivityDate 
\\[1cm]
{\small \VolumeSerial} 
}
\date{}
\maketitle
\thispagestyle{empty}
\end{titlepage}

\baselineskip=14pt
\newpage
\thispagestyle{empty}


\begin{abstract}
Detecting binary black holes in interferometer data requires 
an accurate knowledge of the orbital phase evolution of the
system. From the point of view of data analysis one also
needs fast algorithms to compute the templates that will
employed in searching for black hole binaries. Recently,
there has been progress on both these fronts: On the one
hand, re-summation techniques have made it possible to accelerate
the convergence of poorly convergent asymptotic post-Newtonian series
and derive waveforms beyond the conventional adiabatic approximation.
We now have a waveform model that extends beyond the inspiral
regime into the plunge phase followed by the quasi-normal mode ringing.
On the other hand, explicit Fourier domain waveforms have been derived
that make the generation of waveforms fast enough so as not to be a
burden on the computational resources required in filtering the
detector data. These new developments should make it possible to
efficiently and reliably search for black hole binaries in data
from first interferometers.

\end{abstract}

\vspace{6cm}




\newpage
\thispagestyle{empty}
\tableofcontents

\newpage
\setcounter{page}{1}

\section{Introduction}
One of the theoretical challenges faced in the analysis 
of data from gravitational wave (GW) detectors is the construction
of an efficient and reliable set of templates to search for
binary black holes (BH) and neutron stars (NS). Firstly,
a fast algorithm is needed to compute the waveforms as the parameter
space of compact binaries is quite large requiring several 100,000
templates to cover an interesting range of systems, making it 
prohibitively expensive to store the templates digitally. Secondly,
the templates we employ in our search must be sufficiently accurate
representations of the true GW signal from a BH binary so that we 
miss out, as a result of the mis-match, only a small fraction 
of all possible events. In this talk we will report and discuss
some recent progress made on both these fronts: We now have algorithms
to compute templates, both in the time- and frequency-domains, that take
only one-half of the computational time to filter out the detector
data through these templates. In the time-domain this is achieved by
solving a pair of ordinary differential equations (ODEs) \cite{dis3}, rather
than using parametric representations of the phasing formulas that require
solving integrals of rational polynomials \cite{dis1}. 
In the frequency-domain we have derived explicit analytic expressions 
of the Fourier components \cite{dis2} that
constitute an accurate representation of the time-domain 
signal truncated at, or slightly prior to, the last stable orbit (LSO).
In addition, to an explicit Fourier domain phasing formula we 
also have a pair of ODEs in the frequency-domain
giving the Fourier phase as a function of frequency \cite{dis3}.

In the test mass approximation, the LSO occurs at a frequency $f_{\rm LSO}
= 220 (20M_\odot/M)$~Hz, while for comparable masses it is most
likely to occur at higher frequencies (see, for example Ref. \cite{dis1}).
Until recently, time-domain template waveforms were truncated at the
LSO since we were totally unaware of how to continue the waveform
beyond the last stable orbit. This time truncation, harmless as it may
sound, does mean a specific modelling of the frequency content of 
the merger signal which may or may not be detrimental to extracting the 
true GW signal, depending on how the true signal behaves. Our experiments
showed that if the merger signal terminates quickly, say over less
than half an orbital time-scale, then there is no harm in using the
time-truncated waveforms, or their Fourier-domain counterparts \cite{dis2},
as search templates. However, recent theoretical progress in the Hamiltonian 
description of \cite{djs}, and the effective one-body approach to solving 
the binary black hole dynamics has shed some light on how a binary may
end its life after crossing the LSO \cite{bd00}. More precisely, one has now a 
reasonably good picture of the last three milli-seconds in the life of
a stellar mass static black hole binary when two Schwarzschild black 
holes merge leading to a single spinning Kerr hole. It has been suggested 
that a good strategy to detect compact binaries is to use as search
templates the waveforms based on the effective one-body approach. These
waveforms comprise the inspiral and merger
phases followed by the quasi-normal mode ringing of the newly formed 
spinning black hole.  Such a template bank should be augmented 
by an extended set of search templates, that result from a slight 
variation of the effective one-body waveforms, so as to take into 
account the uncertainty that lies in 
the modelling of the late stages in the evolution of these systems. 

A combination of the sensitivity of initial interferometers and 
the expected compact binary merger rates in the Universe has 
meant that the first likely binary source, an interferometer network 
is likely to observe, is the merger of stellar mass black holes
\cite{postnov et al,dis2,dis3}.  The LIGO-VIRGO-GEO interferometer
network has its peak sensitivity in the frequency interval 
150-400 Hz, which corresponds to a range 11-30 $M_\odot$ of 
the total mass of a system, whose LSO, and hence the most dominant
part of the inspiral signal, occurs in the least noisy band-width of 
the network. Thus, these are our best candidate sources. Fortunately,
the number of templates needed to search for binaries in this range
of masses is less than 10. Hence we will not increase the computational
cost, nor will we worsen the statistics (false alarm/dismissal rates)
by enhancing the number of templates in this range of masses.

\section{Post-Newtonian waveform}

The post-Newtonian (PN) equations of motion and wave generation formalisms,
when applied to a compact binary evolution, facilitate the computation
of the two polarizations $h_+$ and $h_\times$ of the gravitational wave
emitted by the system. For a system in circular orbit, located at a 
distance $r$ from the Earth, the two polarisations are given as
post-Newtonian expansions in the invariant velocity $v$ 
by the following expressions\cite{bdiww,biww}
\begin{equation}
h_{+,\times} = \frac{2 M\eta v^2}{r}
   \left [ H^{(0)}_{+,\times} + \frac{\delta M}{M} H^{(1)}_{+,\times}v 
   + H^{(2)}_{+,\times}v^2 
   + \frac{\delta M}{M} H^{(3)}_{+,\times} v^3 + H^{(4)}_{+,\times}v^4 
    \right ]\ , 
\end{equation}
where
\begin{equation}
H^{(0)}_+ = -(1+\cos^2 {i}) \cos 2\varphi, \ \ \ \ 
H^{(0)}_\times = -2\cos {i} \sin 2\varphi, \ \ \ \ \ldots
\end{equation}
$M=m_1+m_2$ and $\eta= m_1m_2/M^2$ are the total mass and
the symmetric mass ratio of the system, respectively, $\delta M = m_1-m_2,$ $i$
is the inclination of the plane of the binary with respect to the 
line-of-sight and $\varphi$ is the orbital phase whose post-Newtonian
expansion is also known presently to order $v^5.$ For equal mass binaries,
that is, $\delta M = 0$ and $\eta=1/4$, the PN amplitude corrections
$H_{+,\times}^{(n)},$ $n>0,$ are less important than they are for asymmetric 
binaries with $\delta M \sim M$ and $\eta \ll 1.$
Although these {\it amplitude } corrections can be significant in certain
stellar mass binaries,  which the initial ground-based interferometers
are likely to observe, such as a NS-BH binary, it has been customary 
to employ only the dominant term $H_{+,\times}^{(0)}$ in search templates -- 
the so-called {\it restricted} PN approximation \cite{last3min}. 
We shall continue to adopt
this approximation in this paper, but let us note here that a careful
search for asymmetric binaries must include the full PN signal including
all the amplitude corrections \cite{sintes and vecchio}. In the restricted
PN approximation the signal recorded by the detector is the linear
combination 
\begin{equation}
h = F_+h_+ + F_\times h_\times,
\label{eq:wave1}
\end{equation}
where $F_+$ and $F_\times$
denote the antenna patterns of the detector \cite{thorne87}.

The evolution of the orbital 
phase can be worked out by using the PN expansions of the binding
energy per unit mass $E(v)$ of the system and the gravitational wave 
flux $F(v)$ emitted,
in an energy balance equation, namely, $F=-M (dE/dt).$ For two 
bodies of comparable masses in circular orbit one has
\begin{equation}
E(v) = -\frac{\eta v^2}{2}
\left (1 + E_2 v^2 + E_4 v^4 + O(v^6) \right ),
\label{eq:energy}
\end{equation}
\begin{equation}
F(v) = \frac{32\eta^2 v^{10}}{5}
\left (1 + F_2 v^2 + F_3 v^3 + F_4 v^4 + F_5 v^5 + O(v^6)\right ),
\label{eq:flux}
\end{equation}
where 
\begin{equation}
E_2=-\frac{9+\eta}{12}, \ \ 
E_4=-\frac{81-57\eta+\eta^2}{24},
\end{equation}
\begin{equation}
F_2 = - \frac{1247}{336} - \frac{35\eta}{12},\ \ 
F_3 = 4\pi,\ \ 
F_4 = -\frac{44711}{9072} + \frac{9271\eta}{504} + \frac{65\eta^2}{18},\ \ 
F_5 = -\left(\frac{8191}{672} + \frac{535\eta}{24}\right) \pi.
\end{equation}
The phasing of the orbit is given parametrically by the following 
pair of ordinary differential equations (ODEs):
\begin{eqnarray}
\frac{d\varphi}{dt} = \frac{v^3}{M}, \ \ \ \ 
\frac{dv}{dt} = \frac {dv}{dE} \frac{dE}{dt} = -\frac{F(v)}{ME'(v)}
\label {eq:orbital evolution}
\end{eqnarray}
where $E'(v)\equiv dE/dv.$
There are several different, but conceptually equivalent,
possibilities to proceed from the above equations in arriving at an
explicit phasing formula. A straightforward approach is to simply substitute
for $E$ and $F$ their PN expansions, Eqs.~(\ref{eq:energy}) and (\ref{eq:flux}),
re-expand the rational polynomial $F/E'$ to the correct PN order and
solve the above differential equations to yield an explicit
time-domain phasing formula. This is the usual PN approximation
that has also been called the T-approximant \cite{dis1}.
Alternatively, given the PN expansions, or other representations, of the
flux and energy, one can numerically solve the above ODEs 
for equal time steps and then use Eq.~(\ref{eq:wave1}).
It turns out that this latter method is accurate and computationally
quite fast. Other equivalent, but numerically distinct,
approaches are discussed in Ref.~\cite{dis3},
to which we refer the interested reader for an exhaustive account and
comparison. Among the different approaches introduced therein,
we would like to discuss here one in some detail, viz the P-approximants
\cite{dis1,dis2}. P-approximants make the best use of our current
theoretical knowledge and combine that with a fast converging re-summation
technique.  We shall compare its performance against a recently proposed
new class of waveforms, based on a different re-summation technique
\cite{bd00} -- the effective one-body waveforms -- which is likely 
to serve as the most accurate set of search templates.

\section {Effective One-Body Approach}
\label{sec:eob}

As discussed above, in the standard ``adiabatic approximation'' 
the evolution of the orbital phase is constructed by combining
the energy-balance equation $M (dE/dt)=-F$ with either
the PN approximation, or re-summed
estimates, for the energy and flux as 
functions of the instantaneous circular
orbital frequency. Recently, Buonnano and Damour \cite{bd00} introduced
a new approach which is no longer limited
to the adiabatic approximation and expected to describe 
rather accurately the transition between the inspiral and the plunge, 
and to give also an estimate of the ensuing plunge signal. The approach
of \cite{bd00} is essentially, like \cite{dis1,dis2}, a re-summation
technique which consists of two main ingredients: (i) the ``conservative'',
that is non-dissipative, part of the dynamics [effectively equivalent to the
specification of the  $E(v)$ in the previous approaches] 
is re-summed by a an effective one-body dynamics that replaces the
two-body dynamics, and
(ii) the dissipative part of the dynamics [equivalent to the specification of
the  $F(v)$] is constructed by borrowing the re-summation technique 
introduced in \cite{dis1}. In practical terms, the time-domain signal at
the detector is given by the following expression in terms of the reduced 
time $\hat{t}=t/M$:
\begin{equation}
\label{4.1}
 \quad h(\wt) = {\cal
C} \, v_{\omega}^2 (\wt) \cos (2\varphi(\wt))\,, \quad v_{\omega}
\equiv \left( \frac{d \varphi}{d \wt} \right)^{\frac{1}{3}}, 
\end{equation}
where $\cal C$ is a constant for a given binary depending 
on the masses of the two stars, the polarisation of the wave, 
the antenna pattern and the distance to the source.
The orbital phase $\varphi(\wt)$ entering Eq.~(\ref{4.1}) is given by
integrating a system of ODE's:
\begin{eqnarray}
\label{eq:3.28}
&&\frac{dr}{d \wt} = \frac{\pa \wH}{\pa p_r}
(r,p_r,p_\varphi)\,, \\
\label{3.29}
&& \frac{d \varphi}{d \wt} = \ww \equiv \frac{\pa \wH}{\pa p_\varphi}
(r,p_r,p_\varphi)\,, \\
\label{3.30}
&& \frac{d p_r}{d \wt} + \frac{\pa \wH}{\pa r}
(r,p_r,p_\varphi)=0\,, \\
&& \frac{d p_\varphi}{d \wt} = \wF_\varphi(\ww (r,p_r,p_{\varphi}))\,.
\label{3.31}
\end{eqnarray}
The reduced Hamiltonian $\widehat{H}$ (of the associated one-body problem) 
is given, at the 2PN approximation\footnote{The 3PN version of 
$\widehat{H}$ has been recently obtained in Ref. \cite{djs}.}, by 
\begin{eqnarray}
\label{eq:3.32}
\wH(r,p_r,p_\varphi) = \frac{1}{\eta}\,\sqrt{1 + 2\eta\,\left [
\sqrt{A(r)\,\left (1 + \frac{p_r^2}{B(r)} + \frac{p_\varphi^2}{r^2} \right )} -1 \right ]}\,,\\
\label{3.34}
{\rm where}\;\;
A(r) \equiv 1 - \frac{2}{r} + \frac{2\eta}{r^3} \,,
\quad \quad B(r) \equiv \frac{1}{A(r)}\,\left (1 - \frac{6\eta}{r^2}
\right ).
\end{eqnarray}
The damping force ${F}_{\varphi}$ is approximated by
\begin{equation}
\widehat{{ F}}_{\varphi}
=-\frac{1}{\eta v_\omega ^3}{ F}_{P_n}(v_\omega)\,,
\label{eq:damp}
\end{equation}
where $
{ F}_{P_n} (v_{\omega})  
= \frac{32}{5}\,\eta^2\,v_\omega^{10}\,
\hat{{ F}}_{P_n} (v_{\omega})$  
is the flux function used in P-approximants to be discussed below.

The system  of equations (\ref{eq:3.28})--(\ref{3.31}),
allows one to describe the smooth transition which takes place 
between the inspiral and the plunge, while the adiabatic evolution 
becomes spuriously singular at the LSO, 
where $E'(v_{\rm LSO})=0$. Ref.\cite{bd00} advocated to
continue using the above system of equations after the transition, 
to describe the waveform emitted during the plunge
and to match around the ``light ring'' to a ``merger'' 
waveform, described, in the first approximation, 
by the ringing of the least-damped quasi-normal mode
of a Kerr black hole \cite{bd00}.
At the moment this technique is the most complete.
It includes, in the best available 
approximation and for non-spinning black holes, most of the 
correct physics of the problem, and leads to a specific prediction for the
complete waveform (inspiral + plunge + merger) emitted by coalescing binaries.
Because of its completeness, we shall use it as our ``fiducial exact''
waveform in our comparison between different search templates.

The initial data $(r_0, p_{r}^0, p_{\varphi}^0)$ are found using
\begin{equation}
r_0^3 \left [ \frac {1 + 2 \eta (\sqrt{z(r_0)} -1 )}{1- 3\eta/r_0^2} \right ]-  \hat \omega_0^{-2} = 0,\ \ 
p^0_\varphi = \left [\frac {r_0^2 - 3 \eta}{r_0^3 - 3 r_0^2 + 5 \eta} \right ]^{1/2},\ \ 
p^0_r = \frac {{ F}_\varphi(\hat \omega)}{C(r_0,p^0_\varphi) (dp^0_\varphi/dr_0)}\ \ 
\end{equation}
where $z(r)$ and $C(r,p_\varphi)$ are given by
\begin{equation}
z(r) = \frac{r^3 A^2(r)}{r^3-3r^2+5 \eta},\ \ 
C(r,p_\varphi) = \frac{1}{\eta \wH (r,0,p_\varphi)
 \sqrt{z(r)}} \frac{A^2(r)}{(1-6\eta/r^2)}.
\end{equation}
The plunge waveform is terminated when the radial coordinate attains the value
at the light ring $r_{\rm lr}$ given by the solution to the equation,
\begin{equation}
r_{\rm lr}^3 - 3 r_{\rm lr}^2 + 5 \eta = 0.
\end{equation}
The subsequent ``merger'' waveform is constructed as in Ref.\cite{bd00}.
\begin{figure}
\centerline {\epsfxsize 3.2 true in  \epsfbox {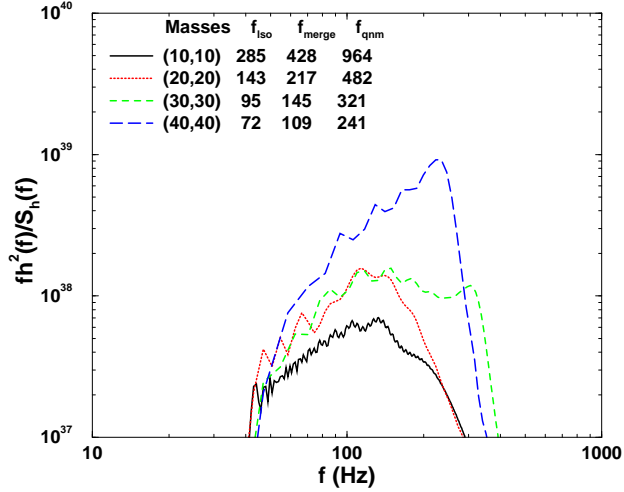} }
\caption{Power spectral density of the signal weighted down 
by the noise spectral density for effective one-body waveforms
for different binary sources.}
\label{fig:signal spectrum}
\end{figure}

In Fig.~\ref{fig:signal spectrum} we show the power spectrum of the 
effective one-body waveform weighted by the noise power spectral
density of LIGO for four different systems. For binaries of
mass less than about 30 $M_\odot$ the signal exhibits mostly the 
inspiral part, for  $30 M_\odot < M < 50 M_\odot$ we see the
inspiral and plunge while for $M>50 M_\odot$ one sees the plunge
and the quasi-normal mode ringing, the inspiral part being 
insignificant.

\section {P-approximants}
\begin{figure}
\centerline {
\epsfxsize 3.2 true in  \epsfbox {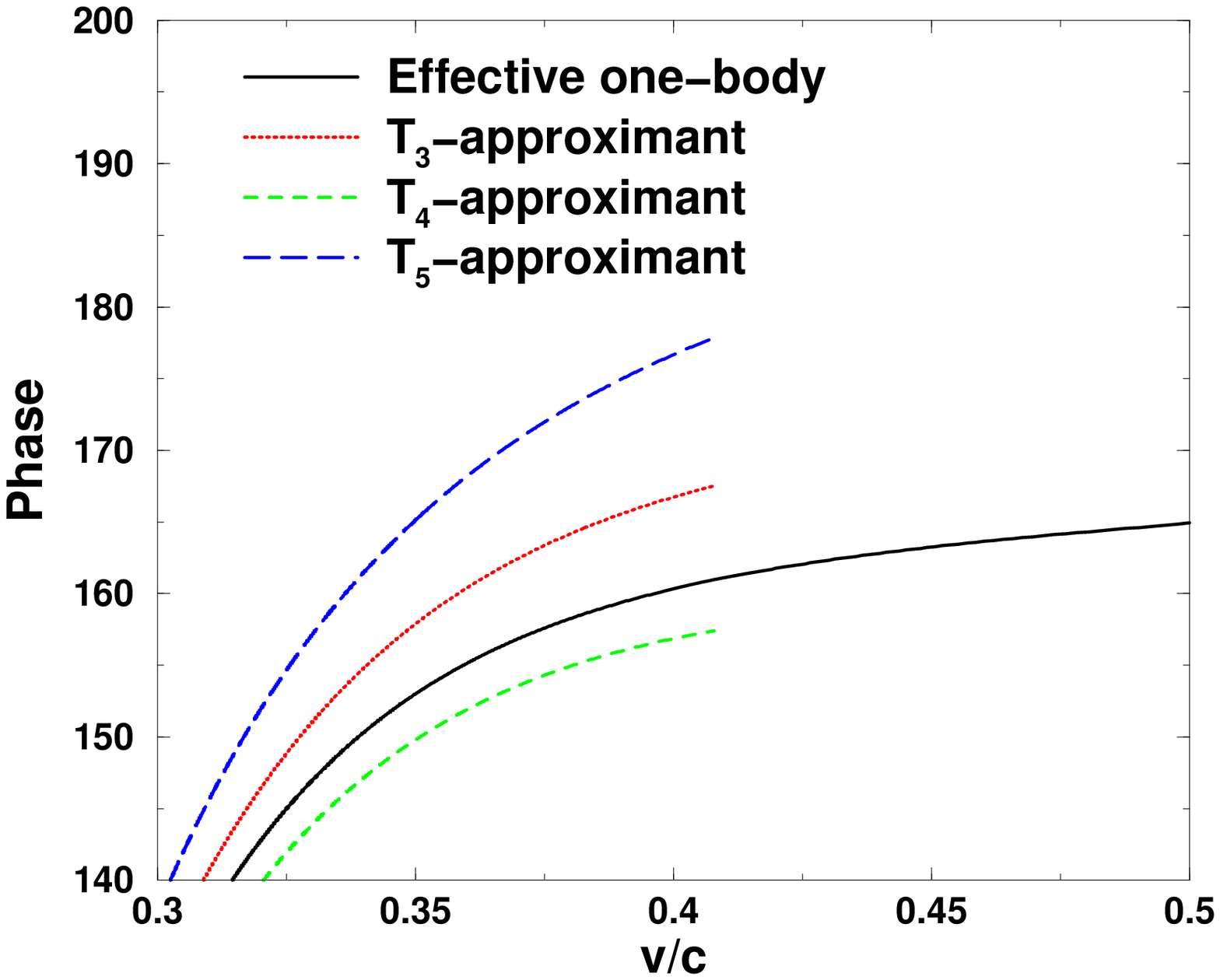} 
\epsfxsize 3.2 true in  \epsfbox {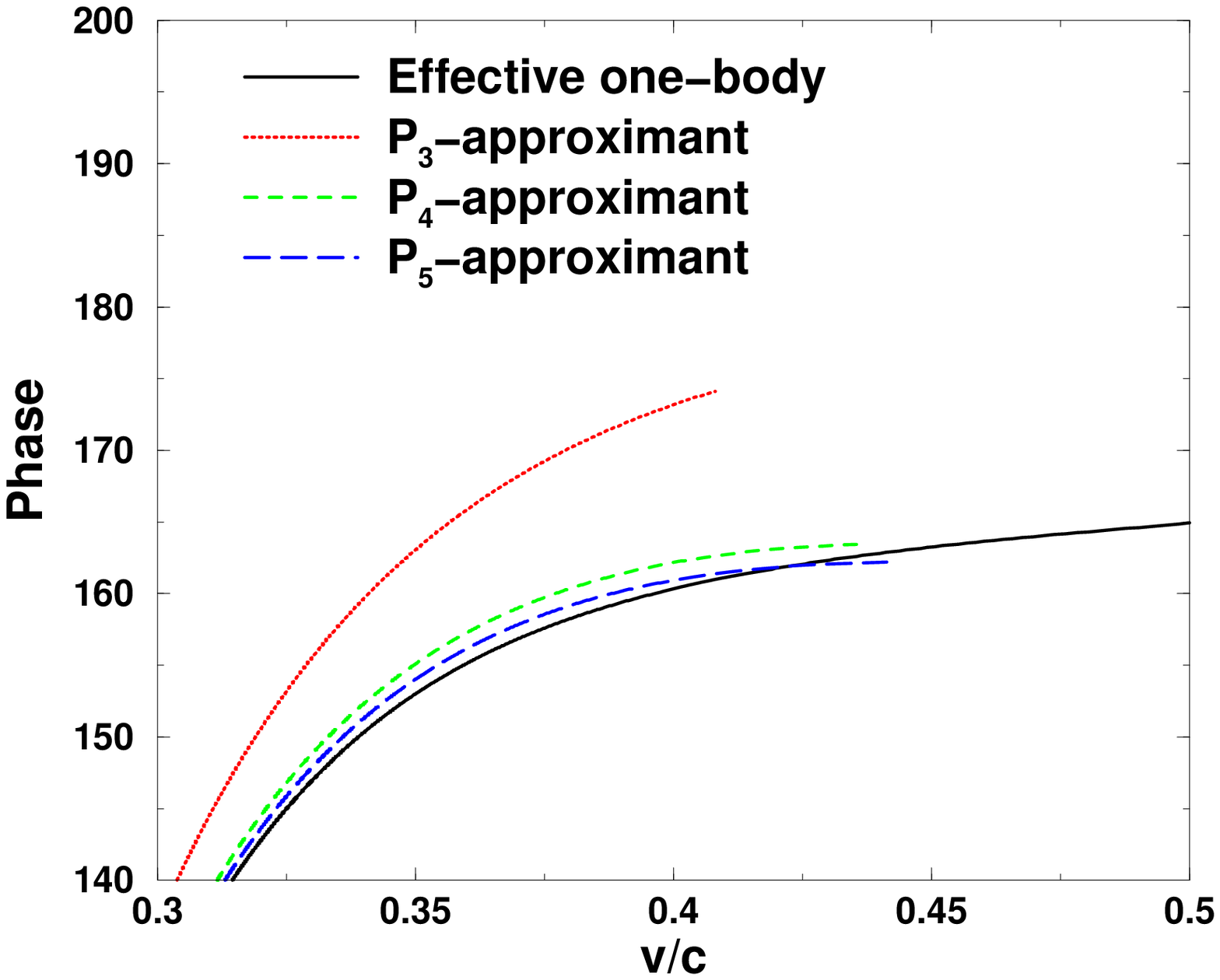} }
\caption{Orbital evolution of a binary consisting of two stars of
equal masses is shown plotted at different post-Newtonian orders.
We show the usual Taylor approximant-based phase evolution on the 
left and the re-summed P-approximant-based phase evolution on the right.}
\label{fig:orbital evolution}
\end{figure}

The PN series for the flux and energy functions, Eqs.~(\ref{eq:energy})
and (\ref{eq:flux}),  are asymptotic series
that converge only in the limit $v\rightarrow 0.$ In our application,
however, we will be using these series in a region where $v \sim 1.$
It is no surprise, therefore, that one finds that the orbital evolution
inferred by solving the ODEs in Eq.~(\ref{eq:orbital evolution}) differ
greatly at different PN orders. More precisely, the orbital evolution
inferred by using the post-Newtonian expansions of $E$ and $F$ in
Eq.~(\ref{eq:orbital evolution}) are significantly different even at high
PN orders when $v$ is close to 1 or when the binary system is close
to coalescence. This is shown in Fig.~\ref{fig:orbital evolution} where
we have plotted $\varphi(v)$ as a function of $v$ at different PN
orders. Also plotted in thick solid line is the orbital evolution
predicted by the effective one-body approach discussed in the
previous Section, which is, presently,
our best guess on how the phase may evolve.  
In the limiting case of a test mass orbiting a Schwarzschild black
hole the post-Newtonian expansion of the phase can be worked out
up to order $O(v^{11}).$ Even at such high PN orders the expansion
is poorly convergent \cite{poisson95,dis1}. 
This poor convergence means that the PN expansions of the phase
are not very reliable and their use in constructing templates will result
in missing up to one in every three events that a detector can
observe with the aid of a more accurate template \cite{dis1}. 
More importantly, the measurement of physical parameters will 
be severely biased, in some cases by as much as 100 \%, thereby making 
gravitational astronomy imprecise. 

The situation can be dramatically improved  by using re-summation 
techniques, which work with quantities that are 
conceptually equivalent to the PN 
expansions, but numerically different from them. These re-summed
quantities are often rapidly convergent because they can capture
a certain physical effect which a PN expansion is never able to.
For instance, on very general grounds it is expected that the
PN expansion of the flux of GW emitted by a test particle inspiralling
in Schwarzschild geometry must exhibit a simple-pole singularity
at the location of the light-right \cite{dis1}, namely $v=1/\sqrt{3}$. A 
straightforward PN expansion of the flux will never be able to
predict this pole while a rational-polynomial approximation to 
the post-Newtonian series, will by construction, yield a pole.

It is often the case in physical problems, that such 
rational-polynomials, called Pad\'e approximants, capture 
the location of the true pole singularity. However, to be able to
employ this technique one must work with the correct physical
quantities so that, to the best of available knowledge, only a
pole singularity occurs in this quantity, and not some other form
of singularity, such as a branch point. To apply the Pad\'e 
techniques successfully, therefore, we have had to begin from 
more `basic' quantities. For instance, instead of working with
the usually defined energy function $E(v),$ which is, incidentally,
asymmetric in the two masses, we defined the energy function $e(v)$
\begin{equation}
e(v) = \left[1 + \frac{1}{2 \eta} \left (E^2+2E \right ) \right]^2-1,
\label{eq:Eofx}
\end{equation}
which we know in the test mass limit possesses only a simple pole 
on the real line and is also symmetric in the two masses.  Indeed, 
as shown in Ref. \cite{dis1}, the rational polynomial constructed 
from just the first two terms in the PN expansion\footnote{We follow 
the convention wherein the Taylor expansion of a quantity $f(v)$ 
to order $v^n$ is denoted as $f_{T_n}(v)$. Thus, 
$f(v) = f_{T_n}(v) + O(v^{n+1})$.}
$e_{T_2}(v)$ reproduces the exact energy $e(v).$ 
Hence, by using, in the case of binaries with two comparable masses, 
this very form energy function as the quantity
on which to apply the Pad\'e technique, we are more likely
to produce the correct pole singularity. As in the case of
energy, we use a new flux function as a starting point 
and use the Pad\'e approximant of the new energy and flux functions
in constructing the phase evolution. We call this approach of introducing
new energy and flux functions and then applying re-summation techniques
as the P-approximant approach.

The P-approximant phase evolution 
is plotted on the right hand panel of Fig.~\ref{fig:orbital
evolution} and again compared to the effective one-body approach. Clearly,
the curves on the right panel are more rapidly convergent and more
closer to the effective one-body phase evolution, than the Taylor approximant
phase evolution on the left panel.

\section {Overlaps}
The visual comparison of the phase evolution discussed in the previous
Section is a qualitative test of the accuracy of different approximation
schemes.  Eventually, our interest lies in the construction of
template waveforms that are effectual in capturing the true signal.
Therefore, of primary concern is the fraction of the optimal 
signal-to-noise ratio (SNR)\footnote{Optimal SNR is that SNR 
which would have been 
achieved had we employed the functional form of the true signal 
as a template in matched filtering the detector output.}
captured by an approximate waveform.  In maximising this fraction
one varies both the extrinsic, that is the initial phase and the
time-of-arrival, as well as the intrinsic, that is the masses
of the component stars, template parameters.  The SNR depends
not only on the properties of the signal but also on the frequency
response of the detector. More precisely, the fraction of the optimal
SNR achievable by a given approximant $\cal O$ is given by
\begin{equation} 
{\cal O} = \max_{\rm parameters} \frac{\left < A, X \right > }
           {\sqrt {\left < A, A \right > \left < X, X \right > } },
\end{equation} 
where, given two functions $A(t)$ and $B(t)$ their scalar 
product $\left<A,B\right >$ is defined as
\begin{equation} 
\left < A, B \right > = 2 \int^\infty_0 \frac{df}{S_h(f)} \tilde A(f)
\tilde B^*(f) + {\rm C.C.}
\end{equation} 
Here $\tilde A(f)$ denotes the Fourier transform of $A(t)$ 
and $\tilde A^*(f)$ is the complex conjugate of $\tilde A(f)$
and $S_h(f)$ denotes the one-sided noise power spectral density of
the detector.

\begin{figure}
\centerline {\epsfxsize 3.2 true in  \epsfbox {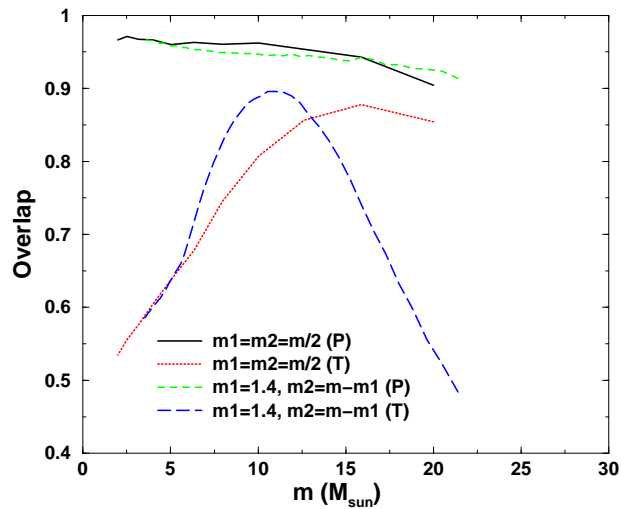} }
\caption{The overlap of T- and P-approximant waveforms
with effective one-body waveforms.}
\label{fig:overlaps}
\end{figure}

In Fig.~\ref{fig:overlaps} we have plotted the overlap of T- and
P-approximants with the effective one-body waveforms as a function 
of the total mass, for equal mass binaries, and for a system in
which one of the bodies is a neutron star. 
We have maximised the inner product only over extrinsic parameters in order
to exhibit how faithful are the two approximants in reproducing
the effective one-body waveform. The superiority of P-approximants
over T-approximants is quite clearly brought about in this plot.
For binaries whose merger takes place at frequencies beyond the
sensitive bandwidth of a detector the disagreement between P-approximants
and the effective one-body approach is 
not too great. However, for sources that merge
in the frequency band of the detector, it is best to employ effective
one-body waveforms as search templates.

Having demonstrated that the two re-summed techniques give very similar
waveforms we suggest that that the effective one-body waveforms be used
as search templates. Indeed, a good strategy would be to employ a
set of search templates based on, but encompassing a larger space than,
the waveforms derived from the effective one-body approach.

\end{document}